\documentclass[12pt,onecolumn]{IEEEtran}

\usepackage[utf8]{inputenc} 
\usepackage[T1]{fontenc}
\usepackage{url}              % provides \url{...}
\usepackage{cite}             % improves presentation of citations

\usepackage[cmex10]{amsmath}  % Use the [cmex10] option to ensure complicance
\usepackage{mleftright}       % fix to wrong spacing of \left-,
\mleftright                   % \middle- \right-commands 

\usepackage{graphicx}         % provides \includegraphics{...} to
\usepackage{booktabs}         % fixes poor spacing in tables and
\usepackage{float}
\usepackage{caption}
\usepackage[utf8]{inputenc} 
\usepackage[T1]{fontenc}
\usepackage{url}
\usepackage{ifthen}
\usepackage{graphicx}
\usepackage{amsthm}
\usepackage{amsfonts}
\usepackage{cite}
\usepackage{multicol}
\usepackage{color}
\usepackage{amssymb}
\usepackage[cmex10]{amsmath} % Use the [cmex10] option to ensure complicance
\usepackage{stfloats}
\usepackage{enumitem}
\newlist{steps}{enumerate}{1}
\setlist[steps, 1]{label = Step \arabic*:}

\newcommand{\un}[1]{\underline{#1}}
\newcommand{\bb}[1]{\mathbb{#1}}

\newcommand{\taust}{\tau^*}
\newtheorem{defn}{Definition}
\newtheorem{thm}{{\cal T}heorem}
\newtheorem{cor}{Corollary}
\newtheorem{prop}{Proposition}
\newtheorem{lem}{Lemma}
\newtheorem{conj}{Conjecture}
\newtheorem{constr}{Construction}
\newtheorem{note}{Note}

\newtheorem{remark}{Remark}
\newtheorem{observation}{Observation}
\newtheorem{argument}{Argument}
\newcommand{\bit}{\begin{itemize}}
	\newcommand{\eit}{\end{itemize}}
\newcommand{\bcor}{\begin{cor}}
	\newcommand{\ecor}{\end{cor}}
\newcommand{\beq}{\begin{equation}}
	\newcommand{\eeq}{\end{equation}}
\newcommand{\beqn}{\begin{equation}}
	\newcommand{\eeqn}{\end{equation}}
\newcommand{\bea}{\begin{eqnarray}}
	\newcommand{\eea}{\end{eqnarray}}
\newcommand{\bean}{\begin{eqnarray*}}
	\newcommand{\eean}{\end{eqnarray*}}
\newcommand{\ben}{\begin{enumerate}}
	\newcommand{\een}{\end{enumerate}}
\newcommand{\bdefn}{\begin{defn}}
	\newcommand{\edefn}{\end{defn}}
\newcommand{\bnote}{\begin{note}}
	\newcommand{\enote}{\end{note}}
\newcommand{\bprop}{\begin{prop}}
	\newcommand{\eprop}{\end{prop}}
\newcommand{\blem}{\begin{lem}}
	\newcommand{\elem}{\end{lem}}
\newcommand{\bthm}{\begin{thm}}
	\newcommand{\ethm}{\end{thm}}
\newcommand{\bobs}{\begin{observation}}
	\newcommand{\eobs}{\end{observation}}
\newcommand{\bconj}{\begin{conj}}
	\newcommand{\econj}{\end{conj}}
\newcommand{\bconstr}{\begin{constr}}
	\newcommand{\econstr}{\end{constr}}
\newcommand{\bpf}{\begin{proof}}
	\newcommand{\epf}{\end{proof}}
\newcommand{\bremark}{\begin{remark}}
	\newcommand{\eremark}{\end{remark}}
\newcommand{\barg}{\begin{argument}}
	\newcommand{\earg}{\end{argument}}
\newcommand{\bprf}{{\em Proof. }}
\newcommand{\eprf}{\hfill $\Box$}
\newcommand{\aw}{(a,w)}
\newcommand{\awe}{(a,w)}
\newcommand{\scerr}{\text{SC}_{\text{\tiny ERR}}}
\newcommand{\swerr}{\text{SW}_{\text{\tiny ERR}}}
\newcommand{\mbswerr}{\text{MBSW}_{\text{\tiny ERR}}}

\newcommand{\R}{R_{\textnormal{opt}}}

\newcommand{\cc}{\mathcal{C}}
	\newcommand{\ccs}{\mathcal{C}_{\text{Str}}}

\newcommand{\baln}{\begin{align*}}
	\newcommand{\ealn}{\end{align*}}
\newcommand{\bal}{\begin{align}}
	\newcommand{\eal}{\end{align}}
\IEEEoverridecommandlockouts

\begin{document}
	
	\title{On Streaming Codes for Burst and Random Errors} 
	
	%%%%%%
	\author{%
		\IEEEauthorblockN{Shobhit Bhatnagar and P. Vijay Kumar}
		\IEEEauthorblockA{%
			\\
			Department of Electrical Communication Engineering, IISc Bangalore\\
			\{shobhitb97,pvk1729\}@gmail.com\\
			\thanks{This research is supported by the "Next Generation Wireless Research and Standardization on 5G and Beyond" project funded by MeitY, SERB Grant No. CRG/2021/008479 and a Qualcomm Innovation Fellowship India 2021.}
	}}
	
	\maketitle
	
	%%%%%
	%% Abstract: 
	%% If your paper is eligible for the student paper award, please add
	%% the comment "THIS PAPER IS ELIGIBLE FOR THE STUDENT PAPER
	%% AWARD." as a first line in the abstract. 
	%% For the final version of the accepted paper, please do not forget
	%% to remove this comment!
	%%
	\begin{abstract}
		Streaming codes (SCs) are packet-level codes that recover erased packets within a strict decoding-delay deadline. Streaming codes for various packet erasure channel models such as sliding-window (SW) channel models that admit random or burst erasures in any SW of a fixed length have been studied in the literature, and the optimal rate as well as rate-optimal code constructions of SCs over such channel models are known. In this paper, we study error-correcting streaming codes ($\text{SC}_{\text{\tiny ERR}}$s), i.e., packet-level codes which recover erroneous packets within a delay constraint. We study $\text{SC}_{\text{\tiny ERR}}$s for two classes of SW channel models, one that admits random packet errors, and another that admits multiple bursts of packet errors, in any SW of a fixed length. For the case of random packet errors, we establish the equivalence of an $\scerr$ and a corresponding SC that recovers from random packet erasures, thus determining the optimal rate of an $\scerr$ for this setting, and providing a rate-optimal code construction for all parameters. We then focus on SCs that recover from multiple erasure bursts and derive a rate-upper-bound for such SCs. We show the necessity of a divisibility constraint for the existence of an SC constructed by the popular diagonal embedding technique, that achieves this rate-bound under a stringent delay requirement. We then show that a construction known in the literature achieves this rate-bound when the divisibility constraint is met. We further show the equivalence of the SCs considered and $\scerr$s for the setting of multiple error bursts, under a stringent delay requirement.
		%, through which we establish the optimal rate and provide rate-optimal code constructions when the parameters of the SW channel satisfy the corresponding divisibility constraint.
	\end{abstract}

	\section{Introduction}
	Erasure-correcting streaming codes, or simply streaming codes (abbreviated as SCs) as they are called in the literature, are packet-level FEC codes that recover erased packets within a strict decoding-delay constraint. They are thus relevant to the current emphasis on low-latency communication. Streaming codes were first studied by Martinian et al. \cite{MartSunTIT04,MartTrotISIT07} for the setting of a sliding-window (SW) channel model that allowed a single erasure burst within any sliding-window of a fixed length. They also introduced the technique of diagonal embedding (DE) which allows one to construct packet-level codes from scalar block codes. This work was then extended in \cite{BadrPatilKhistiTIT17} where the authors considered a SW channel model that allowed either a single erasure burst, or else a few random erasures, in any sliding-window. The optimal rate and rate-optimal code constructions employing DE for this channel model are known for all valid parameters \cite{NikDeepPVK,KhistiExplicitCode}. Since scalar block codes can be used to construct packet-level codes via DE, authors in \cite{LiKhistiGirod} studied the rate-delay trade-off for a class of linear block codes that recover from multiple erasure bursts within a delay constraint. They also presented a code construction that achieves this trade-off.
	We refer the reader to \cite{RamBhaPVK_near_optimal_isit,RudRas_learning_augmented,Khisti_partialRecovery_burst_and_random,ShoBisPVK_burst_and_random,CloMed_multipath,Khisti_three_node_relay,RudRas_online_vs_offline,AdlCas,HagKriKhi_unequal, RLSC2,KriFacDom_3node, RamVajPVK_locally_recoverable_SC,RudRas_learning,Fra_delay_opt,Fra_wt,BadLuiKhi_burst_multicast,MahBadKhis_burst_rank_loss,VajRamNikPVK,VajRamNikPVK_simple_streaming_codes} for the study of streaming codes for various settings and channel models.
	
	In this paper, we study streaming codes for errors, i.e., packet-level codes that recover erroneous packets within a decoding-delay deadline. We will refer to such streaming codes as error-correcting streaming codes (abbreviated as $\scerr$s).
	
	{\emph {Our Contribution}}: We study $\scerr$s for burst and random errors. We establish the equivalence of an $\scerr$ for a SW channel model that allows $\le a$ random packet errors in any SW of length $w,$ and a corresponding SC for a SW channel model that allows $\le 2a$ random packet erasures in any SW of length $w.$ 
	Through this equivalence, and results known in prior literature, we characterize the maximum rate of an $\scerr$ for this setting and also provide a linear field size, rate-optimal code construction for all parameters. 
	We then focus on SCs that recover from multiple erasure bursts. 
	The code construction that achieves the rate-delay trade-off established in \cite{LiKhistiGirod} is causal only under a divisibility constraint. We show the necessity of this divisibility constraint.  We then derive an upper-bound to the rate of an SC for a SW channel model that allows multiple erasure bursts within any SW of length $w$, and show that under a stringent delay requirement, one can achieve this bound via DE iff the parameters of the SW channel satisfy the corresponding divisibility constraint. Then, again for the case of a stringent delay requirement, we show the equivalence of $\scerr$s for SW channels that admit multiple burst errors in any SW of length $w$, and  SCs for SW channels that admit double the number of burst erasures in any SW of length $w$.
	
	The organization of the paper is as follows. In Section \ref{sec:background} we provide the necessary background on SCs and describe the framework for $\scerr$s. In Section \ref{sec:random_errors} we determine the optimal rate of an $\scerr$ that recovers from random packet errors. We discuss $\scerr$s for multiple bursts of packet errors Section \ref{sec:burst_errors_erasures}.
	
	\section{Background and Related Prior Work}
	\label{sec:background}
	Notation: Given integers $m,n$, we will use $[m:n]$ to denote the set $\{m,m+1,\dots,n\}$. We will use $\un{0}_n$ to denote the zero vector of length $n,$ and $I_n$ to denote the $(n\times n)$ identity matrix.  
	For matrix $A\in\bb{F}^{m\times n}_q$, and $\mathcal{I}\subseteq[0:m-1],~\mathcal{J}\subseteq[0:n-1]$, $A(\mathcal{I},\mathcal{J})$ denotes the sub-matrix of $A$ consisting of the rows indexed by $\mathcal{I}$ and the columns indexed by $\mathcal{J}$. For $i\in[0:m-1]$ and $j\in[0:n-1]$, $A(i,:)$ and $A(:,j)$ denote the $i$-th row and the $j$-th column of $A,$ respectively. Similarly, $A(\mathcal{I},:)$ and $A(:,\mathcal{J})$ denote the sub-matrices of $A$ consisting of the rows indexed by $\mathcal{I}$ and the columns indexed by $\mathcal{J}$, respectively.  We will at times use $A_j$ to denote the $j$-th column of $A$. For a vector $\un{v}=[v_0,v_1,\dots,v_{n-1}]^T\in\bb{F}_q^n,$ we will use $\text{supp}(\un{v})$ to denote the support of $\un{v},$ i.e., $\text{supp}(\un{v})=\{i~|~v_i\ne 0\}$. We will use $w_H(\un{v})$ to denote the Hamming weight of $\un{v},$ i.e., $w_H(\un{v})=|\text{supp}(\un{v})|$.
	
	 We will say that an $[n,k]$ code $\cc$ is systematic if 
	 %$\cc=\{\un{c}^T=\un{u}^T G\ |\ \un{u}\in\bb{F}^k_q\},$ where 
	 its generator matrix $G$ is of the form $G=[I_k\ |\ P]\in\bb{F}^{k\times n}_q$. We will call a parity-check matrix $H$ of a systematic code to be in systematic form if it is of the form $H=[P'\ |\ I_{n-k}]\in\bb{F}^{(n-k)\times n}_q$. An $[n,k]$ code $\cc$ is said to be causal if its generator matrix $G$ is of the form $G=[U_{k\times k}~|~ P]\in\bb{F}^{k\times n}_q,$ where $U$ is upper-triangular. Clearly, a systematic code is always causal.
	
	In the context of an $[n,k]$ block code $\cc$, an erasure pattern $\un{e}=[e_0,e_1,\dots,e_{n-1}]^T\in\{0,1\}^n$ signifies that $c_i,~i\in \text{supp}(\un{e})$ is erased, and $c_i,~i\in [0:n-1]\setminus \text{supp}(\un{e})$ is not erased, from a codeword $\un{c}^T=[c_0,c_1,\dots,c_{n-1}]\in\cc.$ %Alternatively, we will sometimes refer to the erasure pattern $\un{e}$ as being described by $\text{supp}(\un{e})$.
	Let $\un{{c}}^T=[{c}_0,{c}_1,\dots,{c}_{n-1}]$ be the codeword corresponding to the message vector $\un{{u}}=[{u}_0,{u}_1,\dots,{u}_{k-1}]^T$. Then, for $i\in[0:k-1],~u_i$ is said to be recoverable from $\un{e}$ within delay $\tau$ if it can be recovered using the non-erased code symbols in $\{{c}_0,{c}_1,\dots,{c}_{\min\{i+\tau,n-1\}}\}.$ Similarly, for $i\in[0:n-1],~{c}_i$ is said to be recoverable from $\un{e}$ within delay $\tau$ if it can be recovered using the non-erased code symbols in $\{{c}_0,{c}_1,\dots,{c}_{\min\{i+\tau,n-1\}}\}.$
	The code $\cc$ is said to be delay-$\tau$ decodable for $\un{e}$ if for all $\un{u}=[u_0,u_1,\dots,u_{k-1}]^T\in\bb{F}^k_q$ and for all $i\in[0:k-1],$ $u_i$ is recoverable from $\un{e}$ within delay $\tau$. 
	
	Whenever we talk about an $[n,k]$ systematic code $\cc$ being delay-$\tau$ decodable for an erasure pattern(s), where $\tau\in[k:n-1],$ we will use the notation $H^{(i)}\triangleq H([0:\tau-k+i],[0:\tau+i]),~i\in[0:n-\tau-2],$ and $H^{(i)} \triangleq H,~i\in[n-\tau-1:n-1]$, where $H$ is the parity-check matrix of $\cc$ in systematic form.
%	%Let $\mathcal{E}$ be a collection of erasure patterns. Then $\cc$ is said to be feasible for all erasure patterns in $\mathcal{E}$ if it can recover from all erasure patterns in $\mathcal{E}$.
%	For a codeword $\un{{c}}^T=[{c}_0,{c}_1,\dots,{c}_{n-1}]\in\cc,$ ${c}_i,~i\in[0:n-1]$ is said to be recoverable within delay $\tau$ if it can be recovered using the non-erased code symbols in $\{{c}_0,{c}_1,\dots,{c}_{\min\{i+\tau,n-1\}}\}$.
%	%A code $\cc$ is said to be delay-$\tau$ decodable if for all $\un{{c}}^T=[{c}_0,{c}_1,\dots,{c}_{n-1}]\in\cc$ and for all $i\in[0:n-1],$ $c_i$ is recoverable within delay $\tau$. 
%	
%%	For a positive integer $B$, an erasure pattern $E\in[0:n-1]$ is said to be a $B$-burst if for some $i\in[0:n-1]$, $E\subseteq[i:i+b-1],$ where $b\le B$. An erasure pattern is said to be a $(z,B)$-burst if it consists of $\le z$ number of $B$-bursts, where $z\ge 1$ is an integer. An $[n,k]$ block code feasible for any erasure pattern of $(z,B)$-bursts is said to be Singleton-achieving if it has minimum possible redundancy, i.e., $n-k=zB$. 
%	
%%	Let $\cc$ be an $[n,k]$ delay-$\tau$ decodable systematic block code such that $\tau\ge k,$ and let $H$ be the parity-check matrix of $\cc$ in systematic form. Define $H^{(i)}\triangleq H([0:\tau-k+i],[0:\tau+i]),~i\in[0:n-\tau-2],$ and $H^{(i)}\triangleq H,~i\in[n-\tau-1:n-1]$.
	
	\begin{figure}
		\begin{center}
			%\captionsetup{font=footnotesize}
			\scalebox{0.6}{\includegraphics{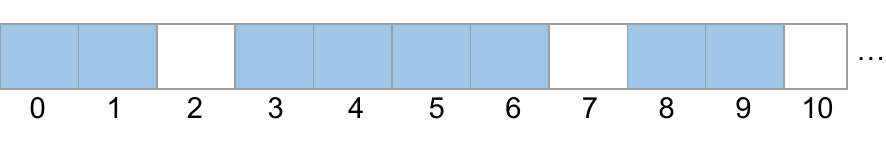}}
			\caption{An admissible erasure pattern of the $(3,2,7)$-MBSW channel. Colored squares denote erasures.}
			\label{fig:multiple_burst_eg}
		\end{center}
		%\vspace{-0.3in} 
	\end{figure}
	
	\subsection{Erasure-Correcting Streaming Codes}
	\paragraph{Framework \cite{MartSunTIT04}} At each time instant $t\in\bb{N}\bigcup\{0\},$ the encoder of a streaming code receives a message packet $\un{u}(t)=[u_1(t),u_2(t),\dots,u_{k}(t)]^T\in\bb{F}^k_q,$ and produces a coded packet $\un{x}(t)=[x_{1}(t),x_2(t),\dots,x_{n}(t)]^T\in\bb{F}^n_q.$ In other words, a message packet contains $k$ symbols from $\bb{F}_q,$ and a coded packet contains $n$ symbols from $\bb{F}_q.$  We will assume that $\un{u}(t)=\un{0}_k,$ for $t<0.$ The streaming code encoder is causal, i.e., $\un{x}(t)$ is a function of only prior message packets $\un{u}(0),\un{u}(1),\dots,\un{u}(t)$. The coded packet $\un{x}(t)$ is then transmitted over a packet erasure channel, so that at time $t,$ the receiver receives 
	$$\un{y}(t)=\begin{cases}
		\un{x}(t),~\text{if}~\un{x}(t) \text{~is not erased,}\\
		\wedge,~\text{if}~\un{x}(t) \text{~is erased.}
	\end{cases}$$
	Here, we use the symbol $\wedge$ to denote an erased packet.
	A delay-constrained decoder at the receiver must recover $\un{u}(t)$ using $\{\un{y}(0),\un{y}(1),\dots,\un{y}(t+\tau)\}$ only, where $\tau$ denotes the decoding-delay constraint, for all $t\ge0$. Packet-level codes satisfying this requirement will be called erasure-correcting streaming codes, or simply, streaming codes (abbreviated as SCs).
	The rate of the SC is defined to be $\frac{k}{n}$. 
	
	In the context of packet-level codes, an erasure pattern $\un{e}=(e_0,e_1,\dots),$ where $e_t\in\{0,1\}$ for all $t,$ denotes that $\un{x}(t)$ is erased if $e_t=1,$ and $\un{x}(t)$ is not erased if $e_t=0$. We will also use the notation $\text{supp}(\un{e})$ to denote the set of time indices of erased packets, i.e., $\text{supp}(\un{e})=\{t~|~e_t=1\}$.

\paragraph{$(a,w)$-SW Channel}
%SW channel
An $\aw$-SW channel is a packet erasure channel that produces only those erasure patterns having the property that any sliding window of $w$ time slots contains $\le a$ erased packets, i.e., $\forall t\ge0,~w_H([e_t,e_{t+1},\dots,e_{t+w-1}])\le a$ for any erasure pattern $\un{e}=(e_0,e_1,\dots)$ of the $\aw$-SW channel. Such erasure patterns will be called admissible erasure patterns of the $\aw$-SW channel.
We will denote a streaming code for the $\aw$-SW channel under delay constraint $\tau$ by an $(a,w,\tau)$-streaming code (abbreviated as $(a,w,\tau)$-SC). Following the arguments in \cite{BadrPatilKhistiTIT17}, one can assume $\tau\ge(w-1)$ w.o.l.o.g., and that an $(a,w,\tau)$-SC of non-zero rate can be constructed iff $a<w$, which we will assume throughout to be the case. The optimal rate of an $(a,w,\tau)$-SC is given by $\frac{w-a}{w}$ \cite{BadrPatilKhistiTIT17,NikDeepPVK}.
%\bea
%\frac{w-a}{w}.
%\label{eq:random_erasure_Ropt}
%\eea	
Further, the streaming code constructed via diagonal embedding (described later) of a $[w,w-a]$ MDS code results in a linear field size, rate-optimal $(a,w,\tau=w-1)$-SC for all valid parameters \cite{BadrPatilKhistiTIT17,NikDeepPVK}.

\paragraph{Multiple Burst-Erasure Sliding Window Channel}
By a $(z,b,w)$-Multiple Burst-Erasure Sliding-Window channel (abbreviated as $(z,b,w)$-MBSW channel), we will mean a packet erasure channel that allows only those erasure patterns having the property that the pattern observed in any sliding window of length $w$ can be described as a $(z,b)$-burst, by which we mean that the pattern observed in any sliding window of length $w$ can be described as consisting of $\le z$ (non-overlapping) erasure bursts, each of length $\le b$. Erasure patterns which satisfy this requirement will be called admissible erasure patterns of the $(z,b,w)$-MBSW channel (see Fig. \ref{fig:multiple_burst_eg} for an example). We remark here that the definition of burst erasures considered in this work does not restrict a burst erasure to necessarily consist only of consecutive erasures, i.e., there can be unerased packets within a burst.

We will denote a streaming code for the $(z,b,w)$-MBSW channel under delay constraint $\tau$ by a $(z,b,w,\tau)$-streaming code (abbreviated as $(z,b,w,\tau)$-SC).
Once again, one can assume w.o.l.o.g. that $\tau\ge (w-1)$. Rate-optimal $(1,b,w,\tau)$-SC constructions are known for all valid parameters \cite{MartSunTIT04,MartTrotISIT07}, and hence we will consider the case $z>1$ throughout the paper.
Further, a streaming code of non-zero rate over the $(z,b,w)$-MBSW channel can be constructed only if $w>zb,$ which we will assume throughout the paper. Finally, we will consider the case $b>1,$ since the case $b=1$ corresponds to random erasures so that the $(z,1,w)$-MBSW channel reduces to the $(z,w)$-SW channel.

Prior work on constructing linear codes that recover from multiple erasure bursts within a delay constraint was done in \cite{LiKhistiGirod}. The authors showed that one can construct an $[n=(k+zb),k]$ code $\cc$ that is delay-$\tau$ decodable for any $(z,b)$-burst iff $\tau\ge\taust\triangleq\max\{k+(z-1)b,zb\}.$ They also provided a code construction that matches this delay bound. However, a close examination of this construction shows that the constructed code is causal only if $b|\taust,$ and further that in this case, one can construct a systematic code. 
%The authors further showed that the delay bound applies to any $[n,k]$ systematic code, even if $n>(k+zB).$ 
The authors also showed, in particular, that it is not possible to construct an $[n=k+zb,k]$ causal block code that is delay-$\taust$ decodable for all $(z,b)$-bursts, when $z>1$ and $k=(b+1).$  In Theorem~\ref{thm:central} below, we will generalize this result to all $k>b$ such that $b\nmid k$.

\paragraph{Diagonal Embedding} In diagonal embedding, codewords of an $[n,k]$ systematic code $\cc$ are placed diagonally in the coded packet stream. See Fig. \ref{fi:DE} for an example.  More formally, if $G=[I_k\ |\ P]$ is a generator matrix of $\cc$, then a stream of $n$ packets in a rate $\frac{k}{n}$ streaming code constructed via DE of $\cc$ is of the form $[x_1(t),x_2(t+1),\dots,x_{n}(t+n-1)]=[u_1(t),u_2(t+1),\dots,u_{k}(t+k-1)]G$. Thus, DE translates packet erasures to codeword symbol erasures. In this paper, we will mostly focus on DE-based code constructions since it is the most popular technique in streaming code literature.

\begin{figure}
	\begin{center}
		\resizebox{.65 \textwidth}{!} 
		{
			\begin{tabular}{ | c | c | c | c | c | c | c | }
				\hline
				{\color{red}\shortstack{$x_1(t)$}} & {\color{blue} \shortstack{$x_1(t+1)$}}&		{\color{black} \shortstack{$x_1(t+2)$}} & &  &  &  
				\\ \hline
				& {\color{red} \shortstack{$x_2(t+1)$}}  & {\color{blue} \shortstack{$x_2(t+2)$}} & {\color{black} \shortstack{$x_2(t+3)$}}&  &  &
			   \\ \hline
				& & 	{\color{red}  \shortstack{$x_3(t+2)$}} & {\color{blue} \shortstack{$x_3(t+3)$}} & {\color{black} \shortstack{$x_3(t+4)$}}  & &
				\\ \hline
				& & &	{\color{red} \shortstack{$x_4(t+3)$}}& {\color{blue} \shortstack{$x_4(t+4)$}} & {\color{black} \shortstack{$x_4(t+5)$}} &
				\\ \hline
				& & & &	{\color{red} \shortstack{$x_5(t+4)$}}& {\color{blue} \shortstack{$x_5(t+5)$}} & {\color{black} \shortstack{$x_5(t+6)$}}  \\ \hline
			\end{tabular}
		}	
		\caption{DE of a $[5, 2]$ systematic code $\mathcal{C}$. Each column is a coded packet and every diagonal of the same color is a codeword of $\mathcal{C}$.}
		\label{fi:DE}
	\end{center}
	%\vspace{-0.3in}
\end{figure}

\subsection{Error-Correcting Streaming Codes}
\begin{figure}
	\begin{center}
		\scalebox{0.6}{\includegraphics{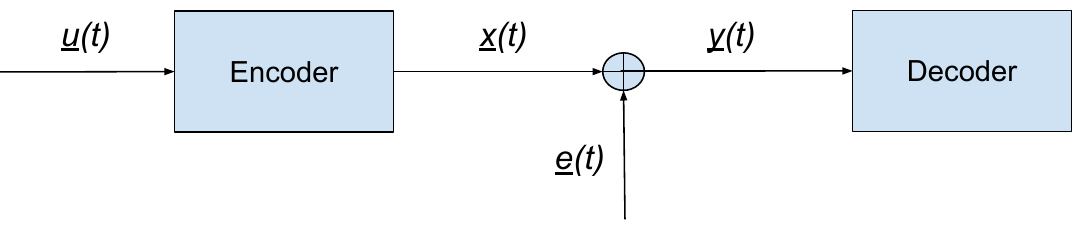}} 
		\caption{Setting of error-correcting streaming codes.}
		\label{fig:error_channel_figure}
	\end{center}
	%\vspace{-0.3in} 
\end{figure}

\paragraph{Framework}
At each time $t\in\bb{N}\bigcup\{0\},$ the encoder receives a message packet $\un{u}(t)=[u_1(t),u_2(t),\dots,u_{k}(t)]^T\in\bb{F}^k_q,$ and produces a coded packet $\un{x}(t)=[x_{1}(t),x_2(t),\dots,x_{n}(t)]^T\in\bb{F}^n_q.$ We will assume that $\un{u}(t)=\un{0}_k,$ for $t<0.$ 
Once again, the encoder is causal, i.e., $\un{x}(t)$ is a function only of prior message packets $\un{u}(0),\un{u}(1),\dots,\un{u}(t)$.
The coded packet $\un{x}(t)$ is then transmitted over a packet error channel, so that at time $t,$ the receiver receives 
$$\un{y}(t)=
\un{x}(t)+\un{e}(t),
$$
where $\un{e}(t)\in\bb{F}_q^n$ (See Fig. \ref{fig:error_channel_figure}).
A delay-constrained decoder at the receiver must recover $\un{u}(t)$ using $\{\un{y}(0),\un{y}(1),\dots,\un{y}(t+\tau)\}$ only, where $\tau$ denotes the decoding-delay constraint, for all $t\ge0$. Packet-level codes that satisfy this requirement will be called error-correcting streaming codes (abbreviated as $\scerr$s). The rate of the $\scerr$ is defined to be $\frac{k}{n}$.

An error pattern $E=(\un{e}(0),\un{e}(1),\dots)$ signifies that $\un{y}(t)=\un{x}(t)+\un{e}(t),~\forall t.$ 
%At times, we will consider error packets in the error pattern $E$ as symbols over $\bb{F}_{q^n},$ and define the vector $\un{V}_E(t)=[\un{e}(t),\un{e}(t+1),\dots, \un{e}(t+\tau)]^T \in \bb{F}^{\tau+1}_{q^n}$. We remark here that even while considering the error packet $\un{e}(t)$ as a symbol over $\bb{F}_{q^n},$ we will retain the underline under the letter $e$, however, it will be clear from context whether we are considering error packets as elements of $\bb{F}_{q^n}$ or $\bb{F}_q^n.$
We will use the notation $\text{supp}(E)$ to denote the set of time indices at which non-trivial packet errors occur, i.e., $\text{supp}(E)=\{t~|~\un{e}(t)\ne0\}$.

\paragraph{$(a,w)$-Error SW Channel}
An $\awe$-Error SW channel (abbreviated as $\awe$-$\swerr$ channel) is a packet error channel that produces only those error patterns having the property that any sliding window of $w$ time slots contains $\le a$ packet errors. 
Alternatively, one can define an $(a,w)$-$\swerr$ channel as a packet error channel that allows only those error patterns $E$ having the property that the erasure pattern $\un{e}=(e_0,e_1,\dots)$ defined as $e_t=1$ if $t\in\text{supp}(E),$ and $e_t=0,$ otherwise, is admissible in the $(a,w)$-SW channel.

We will denote a streaming code for the $\awe$-$\swerr$ channel under delay constraint $\tau$ by an $(a,w,\tau)$-error-correcting streaming code (abbreviated as $(a,w,\tau)$-$\scerr$).
Following similar arguments as in \cite{BadrPatilKhistiTIT17}, one can assume w.o.l.o.g. that $\tau\ge(w-1)$. From the equivalence of an $(a,w,\tau)$-$\scerr$ and a $(2a,w,\tau)$-SC that we will show later in Theorem \ref{thm:random_err_equival}, it follows that one can construct a streaming code of non-zero rate over the $(a,w)$-$\swerr$ channel only if $2a<w,$ which we will assume throughout the paper.
	
\paragraph{Multiple Burst-Error Sliding Window Channel}
By a $(z,b,w)$-Multiple Burst-Error Sliding-Window channel (abbreviated as $(z,b,w)$-$\mbswerr$ channel), we will mean a packet error channel that allows only those error patterns $E$ having the property that the erasure pattern $\un{e}=(e_0,e_1,\dots)$ defined as $e_t=1$ if $t\in\text{supp}(E),$ and $e_t=0,$ otherwise, is admissible in the $(z,b,w)$-MBSW channel.
%the pattern observed in any sliding window of length $W$ can be described as consisting of $\le z$ (non-overlapping) erasure bursts, each of length $\le B$. Erasure patterns which satisfy this requirement will be called admissible erasure patterns of the $(z,B,W)$-MBSW channel (see Fig. \ref{fig:multiple_burst_eg} for an example). 
We will denote a streaming code for the $(z,b,w)$-$\mbswerr$ channel under delay constraint $\tau$ by a $(z,b,w,\tau)$-error-correcting streaming code (abbreviated as $(z,b,w,\tau)$-$\scerr$).
Once again, one can assume w.o.l.o.g. that $\tau\ge (w-1)$. We will also assume $b>1,$ since the case $b=1$ corresponds to random errors so that the $(z,1,w)$-$\mbswerr$ channel reduces to the $(z,w)$-$\swerr$ channel.
%One can argue that $(z,B,W,\tau)$-$\mbswerr$  can be constructed only if $W>zB,$ which we will assume throughout the paper.
% Parameters W>zB

\subsection{Preliminaries}
We will use the following straightforward lemma.
	\blem
	\label{lem:basic}
	Let $\cc$ be an $[n,k]$ code and let $H$ be an $(n-k)\times n$ parity-check matrix of $\cc$. Let the coordinates in $\mathcal{E}\subseteq[0:n-1]$ be erased from $\cc$, and let $i\in\mathcal{E}$. Then, the $i$-th code symbol in any codeword can be recovered from the code symbols corresponding to the coordinates in $[0:n-1]\setminus \mathcal{E}$ iff 
	$$H_i\notin\textnormal{span}\langle\{H_j\ |\ j\in \mathcal{E}\setminus\{i\}\}\rangle.$$
	\elem
	It follows from Lemma \ref{lem:basic} that an $[n,k]$ code $\cc$ can recover from an erasure pattern $\un{e}$ whose support is given by $\mathcal{E}\subseteq[0:n-1]$ iff $\{H_j\ |\ j\in \mathcal{E}\}$ is a linearly independent set, where $H$ is a parity-check matrix of $\cc$. As explained in \cite{NikDeepPVK}, if $\cc$ is a delay-$\tau$ decodable code for an error pattern(s), and $\un{c}^T=[c_0,c_1,\dots,c_{n-1}]\in \cc$, then to recover code symbol $c_i$ such that $i+\tau<n-1$, one must consider the parity-check matrix of the code obtained by puncturing $\cc$ on the coordinates $[i+\tau+1:n-1],$ i.e., the parity-check matrix of the code obtained from $\cc$ be deleting the coordinates $[i+\tau+1:n-1]$ from every codeword of $\cc$. Since the dual of a punctured code is a shortened code, as explained in \cite{NikDeepPVK}, to find a parity-check matrix of the punctured code, we first find a basis for the subspace of the row-space of $H$ which has the property that for every $\un{v}^T=[v_0,v_1,\dots,v_{n-1}]$ in the subspace, $v_\lambda=0,~ \lambda\in[i+\tau+1:n-1]$, and then we delete the coordinates $[i+\tau+1:n-1]$ from every vector in the basis. Specifically, if $\cc$ is a systematic code, and $H$ is its parity-check matrix in systematic form, then the parity-check matrix of the code obtained from $\cc$ by puncturing $\cc$ on the coordinates $[i+\tau+1:n-1]$ is given by $H^{(i)}$, when $\tau\ge k$. Thus, $c_i$ is recoverable form $\un{e}$ within delay $\tau$ iff
	$H^{(i)}_i\notin\textnormal{span}\langle\{H^{(i)}_j\ |\ j\in{\mathcal{E}}\setminus\{i\}\}\rangle.$
	
	Lemmas \ref{lem:full_rank} and \ref{lem:causal_systematic_equivalence} below are proved in the appendix. Lemma \ref{lem:full_rank} below uses Lemma \ref{lem:basic} to identify certain full-rank sub-matrices of the parity-check matrix in systematic form, of an $[n=k+zb,k]$ systematic code that can recover from any $(z,b)$-burst. 
	\blem
	\label{lem:full_rank}
	Let $\cc$ be an $[n=k+zb,k]$ systematic code that can recover from any $(z,b)$-burst, and let $H$ be its parity-check matrix in systematic form. Then, rank$(H([lb:(l+1)b-1],[j:j+b-1]))=b,~l\in[0:(z-1)],~j\in[0:k-b]$.
	\elem
	%	\bprf
	%	Since $H$ is the parity-check matrix of $\cc$ in systematic form, $H(:,[k:n-1])=I_{zB}.$ Consider the erasure pattern $\un{e}$ whose support is given by $\mathcal{E}=[k:n-1]\setminus[lB:(l+1)B-1],$ for some $l\in[0:z-1]$. It is easy to check that $\un{e}$ is a $(z-1)~B$-burst. Also, any vector $\un{v}=[v_0,v_1,\dots,v_{n-1}]^T$ such that $v_\lambda=0,~\lambda\in[lB:(l+1)B-1]$ lies in the span of $\{H_i\mid i\in \mathcal{E}\}$. Now, if for some $j\in[0:k-B],$ rank$(H([lB:(l+1)B-1],[j:j+B-1]))<B$, one can linearly combine $\{H_i\ | \ i\in[j:j+B-1]\}$ to get a vector $\un{v}$ of the form described above. Then, $\{H_i\ |~i\in \mathcal{E}\bigcup\{j:j+B-1\}\}$ becomes a linearly dependent set, and the code can not recover from the $(z,B)$-burst $\un{e}'$ whose support is given by $\mathcal{E}\bigcup\{j:j+B-1\}$.
	%	\eprf
	
	The lemma below establishes the equivalence of a class of causal and systematic codes that are delay-$\tau$ decodable for any $(z,b)$-burst.
	\blem
	\label{lem:causal_systematic_equivalence}
	An $[n=k+zb,k]$ causal block code that is delay-$\tau$ decodable for any $(z,b)$-burst exists iff an $[n=k+zb,k]$ systematic code that is delay-$\tau$ decodable for any $(z,b)$-burst exists.
	\elem	
	
Lemma \ref{lem:main} below will be useful for establishing a divisibility constraint in Theorem \ref{thm:central}.
\blem
\label{lem:main}
Let $b\ge2$ and $m\ge2$ be positive integers. Consider a matrix $A\in\bb{F}^{ 2\times mb}_q$ that satisfies the following properties:
\bit
\item P1: $A(i,[0:b-2])=\un{0}^T_{b-1},$ $i\in\{0,1\}.$ 
\item P2: In any row of $A$, no more than $(b-1)$ consecutive entries can be $0$.
\item P3: For $j\in[b-1:(m-1)b],$ rank$(A(:,[j:j+b-1]))=1$. 
\item P4: For $j\in[0:(m-2)b],$ rank$(A(:,[j:j+2b-1]))=2$.
\eit
Then, $A(i,mb-1)\ne0,\ i\in\{0,1\}$.
\elem

\bprf
From P1 and P2, it follows that $A(i,b-1)\ne0,~i\in\{0,1\}$. Let $A(1,b-1)=\beta\cdot A(0,b-1),$ where$~\beta\in\bb{F}_q\setminus\{0\}$. Suppose that $A(i_1,j_1)\ne 0$ for some $i_1\in\{0,1\}$ and $j_1\in[b:2b-2]$. Then, considering $A(:,[b-1:2b-2]),$ from P3 we have $A(1,j_1)=\beta\cdot A(0,j_1)$. Again, from P3 we have $A(1,[b:2b-1])^T=\beta \cdot A(0,[b:2b-1])^T$. Combining $A(1,[b-1:2b-2])^T=\beta \cdot A(0,[b-1:2b-2])^T$ and $A(1,[b:2b-1])^T=\beta \cdot A(0,[b:2b-1])^T,$ we get $A(1,[0:2b-1])=\beta \cdot A(0,[0:2b-1])^T$, which violates P4. Thus, we have $A(i,[b:2b-2])=\un{0}^T_{b-1},~i\in\{0,1\}$. From P2 we get $A(i,2b-1)\ne0,~i\in\{0,1\}$. Proceeding similarly, the lemma follows.
\eprf
	
	%\section{Rate-Optimal Streaming Codes for Random Errors}
	\section{Optimal Rate of $(a,w,\tau)$-$\scerr$s}
	\label{sec:random_errors}
	In this section, we will drive the optimal rate of an $(a,w,\tau)$-$\scerr$ by establishing the equivalence of an $(a,w,\tau)$-$\scerr$ and a $(2a,w,\tau)$-SC. In the following, at times we will consider error packets in an error pattern as symbols over $\bb{F}_{q^n}\cong\bb{F}_q^n,$ and define the vector $\un{V}_E(t)=[\un{e}(t),\un{e}(t+1),\dots, \un{e}(t+\tau)]^T \in \bb{F}^{\tau+1}_{q^n}$
	 for an admissible error pattern $E=(\un{e}(0),\un{e}(1),\dots)$ of the $(a,w)$-$\swerr$ channel, where $\tau$ is the decoding-delay constraint.
	We remark here that even while considering an error packet $\un{e}(t)$ as a symbol over $\bb{F}_{q^n},$ we will retain the underline under the letter $e$. However, it will be clear from context whether we are considering error packets as elements of $\bb{F}_{q^n}$ or $\bb{F}_q^n.$
	
	The lemma below (proved in the appendix) relates admissible error patterns of the $(a,w)$-$\swerr$ channel to admissible erasure patterns of the $(2a,w)$-SW channel.
	
	\blem
	\label{lem:eq_err_era}
	Let $E=(\un{e}(0),\un{e}(1),\dots)$ and $\tilde{E}=(\un{\tilde{e}}(0),\un{\tilde{e}}(1),\dots)$ be two admissible error patterns of the $(a,w)$-$\swerr$ channel. Let $\un{e}'=(e'_0,e'_1,\dots)$ be the erasure pattern defined as $\un{e}'_t=1$ iff
	$(\un{e}(t)-\un{\tilde{e}}(t))\ne0$. Then $\un{e}'$ is an admissible erasure pattern of the $(2a,w)$-SW channel. Conversely, every admissible erasure pattern of the $(2a,w)$-SW channel can be expressed in this way.
	\elem

	The theorem below establishes the equivalence of an $(a,w,\tau)$-$\scerr$ and a $(2a,w,\tau)$-SC.
	%\label{sec:main}
	\bthm
	An $(a,w,\tau)$-$\scerr$ is a $(2a,w,\tau)$-SC and the converse is also true.
	\label{thm:random_err_equival}
	\ethm
	\bpf
	Let $\cc_{\text{Str}}$ be an $(a,w,\tau)$-$\scerr$. 
	%Since $\ccs$ can recover $\un{u}(t)$ by time slot $(t+\tau)$ for all $t,$ it follows that $\un{x}(t)$ can be recovered by time slot $(t+\tau)$ for all $t$. 
	Let $\{\un{u}(t)\}_{t\ge0}$ be the sequence of message packets transmitted by the source and $\{\un{x}(t)\}_{t\ge0}$ be the corresponding sequence of coded packets.
	In the following, we will consider coded packets $\un{x}(t)$ as scalar symbols over $\bb{F}_{q^n}$, and define the vector $\un{X}{(t)}\triangleq [\un{x}(t),\un{x}(t+1),\dots, \un{x}(t+\tau)]^T \in \bb{F}^{\tau+1}_{q^n}$.
	%Now consider two admissible error patterns $E=(\un{e}(0),\un{e}(1),\dots)$ and $\tilde{E}=(\un{\tilde{e}}(0),\un{\tilde{e}}(1),\dots)$ of the $\awe$-$\swerr$ channel. Consider the erasure pattern $\un{e}'=(e'_0,e'_1,\dots)$ defined as $\un{e}'_t=1$ iff
	%$\un{e}(t)+\un{\tilde{e}}(t)\ne0$. It can be seen that $\un{e}'$ is admissible in the $(2a,w)$-SW channel. 
	Let $\mathcal{E}(a,w)=\{E~\mid\ E~ \text{is admissible in the $\awe$-$\swerr$ channel}\}$ denote the set of admissible error patterns of the $(a,w)$-$\swerr$ channel.

	%$\bb{F}_{q^n},$ for any two admissible error patterns $\un{e}'$ and $\un{\tilde{e}}'$ of the $\awe$-$\swerr$ channel, define for all $t,$ $E'{(t)}\triangleq [\un{e}'(t),\un{e}'(t+1),\dots, \un{e}'(t+\tau)]^T \in \bb{F}^{\tau+1}_{q^n}$ and 
	% $\tilde{E}'{(t)}\triangleq [\un{\tilde{e}}(t),\un{\tilde{e}}(t+1),\dots, \un{\tilde{e}}(t+\tau)]^T \in \bb{F}^{\tau+1}_{q^n}.$ Now consider the erasure pattern $\un{e}=(e_0,e_1,\dots)$ defined as $\un{e}_t=1$ iff
	% $\un{e}'_t+\un{\tilde{e}}'_t\ne0$. It can be verified that $\un{e}$ is an admissible erasure pattern of the $(2a,w)$-SW channel.
	
	We will now show that $\un{u}(0)$ can be recovered from any admissible erasure pattern of the $(2a,w)$-SW channel by time slot $\tau$. Then, one can use induction to prove that $\un{u}(t)$ for any $t>0$ can be recovered from any erasure pattern admissible in the $(2a,w)$-SW channel by time slot $\tau,$ since while recovering $\un{u}(t),$ one can assume that $\{\un{u}(0),\un{u}(1),\dots,\un{u}(t-1)\}$ have been correctly recovered. 
	We remark here that while we keep the proof general, it is sufficient to consider only those error and erasure patterns whose support is a subset of $[0:\tau],$ because while recovering the message packet $\un{u}(0)$ the decoder can only access coded packets received until time slot $\tau.$
%	We remark here that while we keep the proof general, it is sufficient to consider only those error and erasure patterns which non-trivially affect coded packets up to time slot $\tau$ only, i.e., only those error and erasure patterns whose support is a subset of $[0:\tau].$ This is because while recovering the message packet $\un{u}(0),$ the decoder can only access coded packets received until time slot $\tau.$
	
	A necessary and sufficient condition for recovery of $\un{u}(0)$ by time slot $\tau$ is that for any sequence $\{\un{\tilde{x}}(t)\}_{t\ge0}$ of coded packets corresponding to some sequence $\{\un{\tilde{u}}(t)\}_{t\ge0}$ of message packets such that $\un{\tilde{u}}(0)\ne\un{u}(0),$ 
	\begin{align*}
		\mathcal{B}(\un{X}(0)) \bigcap \mathcal{B}(\un{\tilde{X}}(0))=\phi,
	\end{align*}
	where $\un{\tilde{X}}{(t)}\triangleq [\un{\tilde{x}}(t),\un{\tilde{x}}(t+1),\dots, \un{\tilde{x}}(t+\tau)]^T \in \bb{F}^{\tau+1}_{q^n},
	~\mathcal{B}(\un{X}{(0)})=\{\un{X}{(0)}+\un{V}_E(0) ~|~ E\in\mathcal{E}(a,w)\}$ and $\mathcal{B}(\un{\tilde{X}}{(0)})=
	\{\un{\tilde{X}}(0)+\un{V}_{\tilde{E}}(0) ~|~ \tilde{E}\in\mathcal{E}(a,w)\}.$ Equivalently, we must have
	\begin{align}
		\label{eq:x_til-x_1}
		\un{X}(0)-\un{\tilde{X}}(0)\notin \{\un{V}_{\tilde{E}}(0)-\un{V}_E(0)~\mid\ E,\tilde{E}\in\mathcal{E}(a,w)\}.
	\end{align}
	%Now observe the following. Let $E\in\mathcal{E}$ and let $\mathcal{I}=\text{supp}(\un{V}_E(0))$. Then, for any vector $\un{V}\in\bb{F}^{\tau+1}_{q^n}$ such that $\text{supp}(\un{V})=\mathcal{I},$ we have that $\un{V}=\un{V}_{E'}(0)$ for some $E'\in\mathcal{E}$ since non-zero error packets can take any non-zero value in $\bb{F}_q^n$. From this observation, we get that $\eqref{eq:x_til-x_1}$ implies the condition $\text{supp}(X(0)-\tilde{X}(0))\neq \text{supp}(\un{V}_{\tilde{E}}(0)-\un{V}_E(0))$ for any $E,\tilde{E}\in\mathcal{E}$.
	
	Now suppose that there exists an admissible erasure pattern $\un{e}=(e_0,e_1,\dots)$ of the $(2a,w)$-SW channel such that $\ccs$ can not recover $\un{u}(0)$ by time slot $\tau$. Then, it must be the case that there exists a sequence of message packets $\{\un{\tilde{u}}(t)\}_{t\ge0}$ such that $\un{\tilde{u}}(0)\ne\un{u}(0),$ and $\{\un{\tilde{u}}(t)\}_{t\ge0}$ yields the same set of unerased coded packets in the window $[0:\tau]$ when the erasure pattern $\un{e}$ occurs, i.e., $\un{\tilde{x}}(t)=\un{x}(t),~t\in[0:\tau]\setminus\text{supp}([e_0,e_1,\dots,e_\tau])$, where  $\{\un{\tilde{x}}(t)\}_{t\ge0}$ is the sequence of coded packets corresponding to $\{\un{\tilde{u}}(t)\}_{t\ge0}$. Hence, $\text{supp}(	\un{X}(0)-\un{\tilde{X}}(0))\in\text{supp}([e_0,e_1,\dots,e_\tau])$. Using Lemma \ref{lem:eq_err_era}, one can show that there exist error patterns $E,\tilde{E}\in\mathcal{E}(a,w)$ such that $\un{X}(0)-\un{\tilde{X}}(0)= \un{V}_{\tilde{E}}(0)-\un{V}_E(0)$.
	
	Similarly, one can  use \eqref{eq:x_til-x_1} and Lemma \ref{lem:eq_err_era} to show that a $(2a,w,\tau)$-SC is an $(a,w,\tau)$-$\scerr$.
	\epf
	
	The theorem below characterizes the optimal rate of an $(a,w,\tau)$-$\scerr$, and follows from \cite{BadrPatilKhistiTIT17,NikDeepPVK} and Theorem \ref{thm:random_err_equival}.
	\bthm
	The optimal rate of an $(a,w,\tau)$-$\scerr$ is given by $$\R=\frac{w-2a}{w}.$$ Further, DE of a $[w,w-2a]$ MDS code yields a rate-optimal $(a,w,\tau=w-1)$-$\scerr$.
	\label{thm:random_errsc_opt_rate}
	\ethm
	
	Clearly, an $(a,w,\tau=w-1)$-$\scerr$ is also an $(a,w,\tau'>w-1)$-$\scerr$. Thus, the theorem above provides a rate-optimal code construction for all parameters.

\section{Streaming Codes for Multiple Erasure and Error Bursts}
	\label{sec:burst_errors_erasures}
	
	\subsection{Streaming Codes for Multiple Erasure Bursts}
	We will first establish the necessity of a divisibility constraint for the existence of an $[n=k+zb,k]$ causal block code that is delay-$\taust$ decodable for any $(z,b)$-burst. Recall that $\taust=(k+(z-1)b)$ if $k\ge b$.
	\bthm
	\label{thm:central}
	It is possible to construct an $[n=k+zb,k]$ causal block code that is delay-$\taust$ decodable for any $(z,b)$-burst, where $k\ge b,$ iff $b|\taust$.
	\ethm
	\begin{proof} (The proof for an illustrative example case is provided here; the general proof is contained within the appendix.)  
		If $b|\taust,$ one can construct such a code via the construction provided in \cite{LiKhistiGirod}. Therefore, we will focus on necessity here. We will show necessity for an example case.
		From Lemma \ref{lem:causal_systematic_equivalence}, it suffices to show necessity for systematic codes.
		Let $\cc$ be a $[9,5]$ systematic code that is delay-$\taust$ decodable for any $(2,2)$-burst, where $\taust=7.$ Let $\un{c}^T=[c_0,c_1,\dots,c_{8}]$ be an arbitrary codeword of $\cc$, and let $H$ be the parity-check matrix of $\cc$ in systematic form. 
		We will first identify the rank of certain sub-matrices of $H.$ Then, we will invoke Lemma \ref{lem:main} to show that a certain entry of $H$ must be non-zero, using which we will contradict the fact that $\cc$ is delay-$7$ decodable for any $(2,2)$-burst. 
		
		Consider the recovery of code symbol $c_0$. We focus on $H^{(0)}$, which is of size $(3\times8)$. Note that the last 3 columns of $H^{(0)}$ form the $(3\times 3)$ identity matrix, i.e., $H^{(0)}(:,[5:7])=I_3$. Consider the erasure pattern whose support is given by $[0:1]\bigcup[6:7]$. We must have $H^{(0)}(0,0)\ne0$ and $H^{(0)}(0,1)=0,$ because otherwise one could add a (possibly $0$) scalar multiple of $H^{(0)}_1$ to $H^{(0)}_0$ to get a vector $\un{v}=[v_0,v_1,v_2]^T,$ where $v_0=0$, and such a vector lies in the span of $\{H^{(0)}_6,H^{(0)}_7\}$. Similarly, considering the erasure pattern whose support is given by $[0:1]\bigcup[5:6],$ we get that $H^{(0)}(2,0)\ne0$ and $H^{(0)}(2,1)=0.$
		
		Now consider recovery of the symbol $c_1$. Note that $H^{(1)}=H$. Consider the erasure pattern whose support is given by $[1:2]\bigcup[7:8]$. Noting that $H^{(1)}(0,1)=0,$ we get $H^{(1)}(1,1)\ne0$. Similarly, considering the erasure pattern whose support is given by $[1:2]\bigcup[6:7]$, we get $H^{(1)}(3,1)\ne0$. 
		
		Once again consider the recovery of symbol $c_0$. Note that in $H^{(0)}$, $H^{(0)}(0,2)\ne0,$ since from Lemma \ref{lem:full_rank}, $H^{(0)}([0:1],[1:2])$ must be full-rank. Similarly, $H^{(0)}(2,2)\ne0$. Now suppose that $H^{(0)}(\{0,2\},[2:3])$ is full-rank. Then, $H^{(0)}(\{0,2\},0)\in\text{span}\langle\{H^{(0)}(\{0,2\},2),H^{(0)}(\{0,2\},3)\}\rangle$. Since $H^{(0)}(\{0,2\},1)=\un{0}_2^T$ and $H^{(0)}(1,1)\ne0,$ it follows that $H^{(0)}_0\in\text{span}\langle\{H^{(0)}_1,H^{(0)}_2,H^{(0)}_3\}\rangle.$ Thus, rank$(H^{(0)}(\{0,2\},[2:3]))=1$ must hold (rank$(H^{(0)}(\{0,2\},[2:3]))=0$ violates Lemma \ref{lem:full_rank}). Similarly, rank$(H^{(0)}(\{0,2\},[3:4]))=1$ must hold. In the proof for the general case, this argument has been called Argument $1$. We must also have rank$(H^{(0)}(\{0,2\},[1:4]))=2$ from Lemma \ref{lem:basic}. It follows that $H^{(0)}(\{0,2\},[1:4])$ satisfies all the properties of Lemma \ref{lem:main}. Hence we get $H^{(0)}(2,4)\ne0$. However, since $H^{(0)}(\{0,2\},5)=[1,0]^T$, the rank of $H^{(0)}(\{0,2\},\{4,5\})$ is $2$, and the code symbol $c_0$ can not be recovered from the $(2,2)$-burst whose support is given by $[0:1]\bigcup[4:5],$ within delay $7$.
	\end{proof}

	In the following theorem (for proof see appendix), we derive a rate-upper-bound for a $(z,b,w,\tau)$-SC and show that it can be achieved via DE iff a divisibility constraint is met.
	
\begin{figure}
		\begin{center}
			\includegraphics[width=4in]{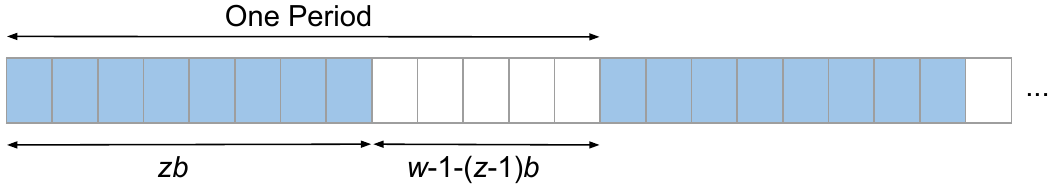} 
			\caption{An admissible periodic erasure pattern of the $(z,b,w)$-MBSW channel.}
			\label{fig:multi_burst_erasure_pattern}
		\end{center}
		%\vspace{-0.3in} 
	\end{figure}

	\bthm
	\label{thm:(z,B,W)-streaming_codes}
	The rate $R$ of a $(z,b,w,\tau)$-SC is upper-bounded as
	\beqn
	\label{eq:mbsw_bound}
	R\le \frac{w-1-(z-1)b}{w-1+b}.
	\eeqn
	Furthermore, a $(z,b,w,\tau=w-1)$-SC that achieves the bound \eqref{eq:mbsw_bound} can be constructed via DE iff $b|(w-1)$.
	\ethm 
	\bprf
	Consider the periodic erasure pattern shown in Fig. \ref{fig:multi_burst_erasure_pattern}. It can be verified that this erasure pattern is admissible in the $(z,b,w)$-MBSW channel. A period of this erasure pattern has length $(w-1+b)$, and in each period there are $(w-1-(z-1)b)$ non-erased packets.  Following similar arguments as in \cite{BadrPatilKhistiTIT17}, it follows that the rate $R$ of a $(z,b,w,\tau)$-SC is upper-bounded as \eqref{eq:mbsw_bound}.
	
	%Substituting $R=\frac{W-1-(z-1)B}{W-1+B}$ in \eqref{eq_Khisti_1}, we get $\tau^\star=W-1$. Thus, if $B|(W-1),$ from Observation 1 we get that DE of the systematic $[W-1+B,W-1-(z-1)B]$ block code constructed via Construction I (for rate $\frac{W-1-(z-1)B}{W-1+B}$) is a streaming code for the $(z,B,W)$-MBSW channel under delay constraint $\tau=W-1$. The rate of this streaming code matches \eqref{eq:mbsw_bound}.
	One can construct a systematic $[w-1+b,w-1-(z-1)b]$ code that is delay-$(w-1)$ decodable for any $(z,b)$-burst using the construction in \cite{LiKhistiGirod} (since $\taust=w-1$ for this case) if $b|(w-1).$ DE of this code yields a $(z,b,w,\tau=w-1)$-SC whose rate matches \eqref{eq:mbsw_bound}.
	
	Now suppose that $\cc$ is an $[n,k]$ systematic code such that DE of $\cc$ results in a $(z,b,w,\tau=w-1)$-SC $\cc_\text{Str}$ whose rate matches \eqref{eq:mbsw_bound}. Since $\cc$ recovers from $zb$ erasures, it follows that $n-k\ge zb$. Since the rate of $\cc_{\text{Str}}$ matches \eqref{eq:mbsw_bound}, we must have $k\ge (w-1-(z-1)b).$ Suppose that a diagonally embedded codeword $\un{c}^T=[c_0,c_1,\dots,c_{n-1}]\in\cc$ is affected by the $(z,b)$-burst whose support is given by $I=[0:b-1]\bigcup[w-(z-1)b:w-1]$. Clearly, if $k>(w-1-(z-1)b),$ there are no parity symbols in $[0:w-1]\setminus I,$ and the code symbol $c_0$ can not be recovered within delay $(w-1)$. Thus, $\cc$ is a $[w-1+b,w-1-(z-1)b]$ systematic code that is delay-$(w-1)$ decodable for any $(z,b)$-burst. From Theorem \ref{thm:central}, it follows that $b|(w-1).$
	\eprf		

As mentioned before, $(1,b,w,\tau=w-1)$-SCs whose rate matches the bound \eqref{eq:mbsw_bound} can be constructed by DE for all parameters \cite{MartSunTIT04,MartTrotISIT07}. Theorem \ref{thm:(z,B,W)-streaming_codes} then in particular shows that a $(1,zb,w,\tau=w-1)$-SC is not necessarily a $(z,b,w,\tau=w-1)$-SC.

\subsection{Streaming Codes for Multiple Error Bursts}
It can be shown that admissible error patterns of the $(z,b,w)$-$\mbswerr$ channel whose support is a subset of $[0:w-1]$ are related to admissible erasure patterns of the $(2z,b,w)$-MBSW channel whose support is a subset of $[0:w-1]$ in a way that is similar to the one described in Lemma \ref{lem:eq_err_era}. It follows that one can construct a $(z,b,w,\tau=w-1)$-$\scerr$ of non-zero rate only if $w>2zb,$ which we will assume in the following.
The theorem below establishes the equivalence of a $(z,b,w,\tau=w-1)$-$\scerr$ and a $(2z,b,w,\tau=w-1)$-SC, and can be proved analogous to Theorem \ref{thm:random_err_equival}.
\bthm
\label{thm:burst_err_equival}
A $(z,b,w,\tau=w-1)$-$\scerr$ is a $(2z,b,w,\tau=w-1)$-SC and the converse is also true.
\ethm

We can now combine Theorem \ref{thm:(z,B,W)-streaming_codes} and Theorem \ref{thm:burst_err_equival} to get the following theorem
analogous to Theorem \ref{thm:random_errsc_opt_rate}.
%about the optimal rate of a $(z,b,w,\tau=w-1)$-$\scerr$.
\bthm
\label{thm:(z,B,W)-Err_streaming_codes}
The rate $R'$ of a $(z,b,w,\tau=w-1)$-$\scerr$ is upper-bounded as
\beqn
\label{eq:err_mbsw_bound}
R'\le \frac{w-1-(2z-1)b}{w-1+b}.
\eeqn
Furthermore, a $(z,b,w,\tau=w-1)$-$\scerr$ that achieves the bound \eqref{eq:err_mbsw_bound} can be constructed via DE iff $b|(w-1)$.
\ethm 	
	
%\newpage
\bibliographystyle{IEEEtran}
\bibliography{Streaming}
%%%%%%%%%%%%%%%%%%%%%%%%%%%%%% 
%\newpage
\appendix
\subsection{Proof of Lemma \ref{lem:full_rank}}
\bprf
Since $H$ is the parity-check matrix of $\cc$ in systematic form, $H(:,[k:n-1])=I_{zb}.$ Consider the erasure pattern $\un{e}$ whose support is given by $\mathcal{E}=[k:n-1]\setminus[lb:(l+1)b-1],$ for some $l\in[0:z-1]$. It is easy to check that $\un{e}$ is a $(z-1,b)$-burst. Also, any vector $\un{v}=[v_0,v_1,\dots,v_{n-1}]^T$ such that $v_\lambda=0,~\lambda\in[lb:(l+1)b-1]$ lies in the span of $\{H_i\mid i\in \mathcal{E}\}$. Now, if for some $j\in[0:k-b],$ rank$(H([lb:(l+1)b-1],[j:j+b-1]))<b$, one can linearly combine $\{H_i\ | \ i\in[j:j+b-1]\}$ to get a vector $\un{v}$ of the form described above. Then, $\{H_i\ |~i\in \mathcal{E}\bigcup\{j:j+b-1\}\}$ becomes a linearly dependent set, and the code can not recover from the $(z,b)$-burst $\un{e}'$ whose support is given by $\mathcal{E}\bigcup\{j:j+b-1\}$.
\eprf
	
\subsection{Proof of Lemma \ref{lem:causal_systematic_equivalence}}
\bprf
An $[n=k+zb,k]$ systematic block code that is delay-$\tau$ decodable for any $(z,b)$-burst is also a causal code, hence sufficiency follows. Now suppose that 
%$\cc=\{\un{c}^T=\un{u}^T G \ | \ \un{u}\in\bb{F}^k_q\},$
$\cc$ is an $[n=k+zb,k]$ causal block code that is delay-$\tau$ decodable for any $(z,b)$-burst. Let $G=[U_{k\times k}\ |\ P]$ be the generator matrix of $\cc$. Since $\cc$ can recover from the $(z,b)$-burst whose support is given by $[n-zb:n-1],~U_{k\times k}$ must be full-rank. Let $\un{c}^T=[c_0,c_1,\dots,c_{n-1}]$ be the codeword corresponding to the message vector $\un{u}=[u_0,u_1,\dots,u_{k-1}]^T$. Since $\cc$ recovers message symbol $u_i,~i\in[0:k-1]$ from any $(z,b)$-burst within delay $\tau$, and $c_j,~j\in[0:k-1]$ is a function of $\{u_0,u_1,\dots,u_j\}$ only (because $U_{k\times k}$ is upper-triangular), it follows that $c_j,~j\in[0:k-1]$ is recoverable within delay $\tau.$
%$U$ is upper-triangular it follows that $c_i,~i\in[0:k-1]$ is recoverable within delay $\tau$ since the message symbols $u_\lambda,~\lambda\in[0:k-1]$ are recoverable within delay $\tau$, and $c_i$ is a function of $u_0,u_1,\dots,u_i$ only. 
Now consider the systematic code %$\hat{\cc}=\{\hat{\un{c}}^T=\hat{\un{u}}^T \hat{G}\ | \ \hat{\un{u}}\in\bb{F}^k_q\}$, 
$\hat{\cc}$ whose generator matrix $\hat{G}=[I_k\ |\ \hat{P}]$ is obtained by performing row operations on $G$. Since the code $\hat{\cc}$ also has the property that for any $\hat{\un{c}}^T=[\hat{c}_0,\hat{c}_1,\dots,\hat{c}_{n-1}]\in\hat{\cc}$, $\hat{c}_j,~j\in[0:k-1]$ is recoverable within delay $\tau$, the claim follows.
\eprf

%\subsection{Proof of Lemma \ref{lem:main}}
%\bprf
%From P1 and P2, it follows that $A(i,b-1)\ne0,~i\in\{0,1\}$. Let $A(1,b-1)=\beta\cdot A(0,b-1),$ where$~\beta\in\bb{F}_q\setminus\{0\}$. Suppose that $A(i_1,j_1)\ne 0$ for some $i_1\in\{0,1\}$ and $j_1\in[b:2b-2]$. Then, considering $A(:,[b-1:2b-2]),$ from P3 we have $A(1,j_1)=\beta\cdot A(0,j_1)$. Again, from P3 we have $A(1,[b:2b-1])^T=\beta \cdot A(0,[b:2b-1])^T$. Combining $A(1,[b-1:2b-2])^T=\beta \cdot A(0,[b-1:2b-2])^T$ and $A(1,[b:2b-1])^T=\beta \cdot A(0,[b:2b-1])^T,$ we get $A(1,[0:2b-1])=\beta \cdot A(0,[0:2b-1])^T$, which violates P4. Thus, we have $A(i,[b:2b-2])=\un{0}^T_{b-1},~i\in\{0,1\}$. From P2 we get $A(i,2b-1)\ne0,~i\in\{0,1\}$. Proceeding similarly, the lemma follows.
%\eprf

\subsection{Proof of Lemma \ref{lem:eq_err_era}}
	\bpf
	Considering the error packets in the admissible error pattern $E$ as symbols over $\bb{F}_{q^n}\cong\bb{F}_q^n,$ we define the vector $\un{U}_E(t)=[\un{e}(t),\un{e}(t+1),\dots, \un{e}(t+w-1)]^T \in \bb{F}^{w}_{q^n}$. Similarly, we define the vector $\un{U}_{\tilde{E}}(t)=[\un{\tilde{e}}(t),\un{\tilde{e}}(t+1),\dots, \un{\tilde{e}}(t+w-1)]^T \in \bb{F}^{w}_{q^n}$ for the error pattern $\tilde{E}.$
	%Considering the error packets in an admissible error pattern $E=(\un{e}(0),\un{e}(1),\dots)$ of the $(a,w)$-$\swerr$ channel as symbols over $\bb{F}_{q^n}\cong\bb{F}_q^n,$ define the vector $\un{U}_E(t)=[\un{e}(t),\un{e}(t+1),\dots, \un{e}(t+w-1)]^T \in \bb{F}^{w}_{q^n}$.
    Now, since $E$ and $\tilde{E}$ are admissible error patterns of the $(a,w)$-$\swerr$ channel, we have $w_H(\un{U}_E(t))\le a$ and $w_H(\un{U}_{\tilde{E}}(t))\le a,~\forall t\ge0.$
    Thus, $w_H([\un{e}'_t,\un{e}'_{t+1},\dots,\un{e}'_{t+w-1}])= w_H(\un{U}_E(t)-\un{U}_{\tilde{E}}(t))\le2a,~\forall t\ge0,$ which implies that $\un{e}'$ is admissible in the $(2a,w)$-SW channel. 

We now prove the converse. Let $\un{e}'$ be an admissible erasure pattern of the $(2a,w)$-SW channel, and let $\mathcal{I}=\{i_0,i_1,\dots\}$ be the (possibly finite) support of $\un{e}'$. Now define two sets $I_1=\{i_j\ |\ j\ \text{is even}\}$ and $I_2=\{i_j\ |\ j\ \text{is odd}\}.$ Note that $I_1\bigcap I_2=\phi$ and $I_1\bigcup I_2=\mathcal{I}.$
Define the error pattern $E=(\un{e}(0),\un{e}(1),\dots)$ as $\un{e}(t)=[1,0,\dots,0]\in\bb{F}_q^n$ (or any other non-zero vector in $\bb{F}_q^n$) if $t\in I_1,$ and $\un{e}(t)=\un{0}_n,$ otherwise. Define the error pattern $\tilde{E}=(\un{\tilde{e}}(0),\un{\tilde{e}}(1),\dots)$ as $\un{\tilde{e}}(t)=[1,0,\dots,0]\in\bb{F}_q^n$ (or any other non-zero vector in $\bb{F}_q^n$) if $t\in I_2,$ and $\un{\tilde{e}}(t)=\un{0}_n,$ otherwise. We will now show that $E$ is admissible in the $(a,w)$-$\swerr$ channel, the proof for $\tilde{E}$ is analogous. Suppose that $E$ is not admissible in the $(a,w)$-$\swerr$ channel. Then, there must exist some $t\ge0$ such that $w_H(\un{U}_E(t))>a.$ In this case, we must have $w_H([\un{e}'_t,\un{e}'_{t+1},\dots,\un{e}'_{t+w-1}])>2a,$ which violates the fact that $\un{e}'$ is an admissible erasure pattern of the $(2a,w)$-SW channel. It is easy to verify that $\un{e}'$ can be obtained from $E$ and $\tilde{E}$ as described in the lemma.
\epf

\subsection{Proof of Theorem \ref{thm:central}}
\bprf
	%We first show sufficiency. Substituting $R=\frac{k}{k+zB}$ in \eqref{eq_Khisti_1}, we get $\tau^\star=(k+(z-1)B)$. Consider the $[\tau^\star+B,\tau^\star-(z-1)B]$ code constructed for rate $R$ via Construction I. Since $B|\tau^\star$, the constructed code is causal, and has dimension $\tau^\star-(z-1)B=k$ and rate $\frac{k}{k+zB}$. 
	%Let $\cc$ be an $[n,k]$ delay-$\tau$ decodable systematic block code such that $\tau\ge k,$ and let $H$ be the parity-check matrix of $\cc$ in systematic form. Define $H^{(i)}\triangleq H([0:\tau-k+i],[0:\tau+i]),~i\in[0:n-\tau-2],$ and $H^{(i)}\triangleq H,~i\in[n-\tau-1:n-1]$.
	We will now show necessity for the general case. From Lemma \ref{lem:causal_systematic_equivalence}, it suffices to show necessity for systematic codes. 
	Let $\cc$ be an $[n=k+zb,k]$ systematic block code that is delay-$\taust$ decodable for any $(z,b)$-burst, where $k\ge b,$ and $\taust=k+(z-1)b.$ It follows that$~b\nmid \taust\iff b\nmid k$. Let $k=ub+a,~a\in[1:b-1]$.
	Let $H$ be the parity-check matrix of $\cc$ in systematic form, and let $\un{c}^T=[c_0,c_1,\dots,c_{n-1}]$ be any arbitrary codeword of $\cc$. Consider the following two cases:
	\paragraph{Case $z=2$}
	We will first show that $H^{(0)}([0:b-1],[0:b-1])$ is lower triangular  with $H^{(0)}(i,i)\ne0,~i\in[0:b-1]$.
	
	Consider recoverability of code symbol $c_0$. Focus on $H^{(0)}$. Note that $H^{(0)}$ is of size $(b+1)\times(n-b+1)$, and the last $(b+1)$ columns of $H^{(0)}$ form an identity matrix. Consider the $(2,b)$-burst whose support is given by $[0:b-1]\bigcup[n-2b:n-b]\setminus\{n-2b\}$. We must have $H^{(0)}(0,0)\ne0$ and $H^{(0)}(0,[1:b-1])=\un{0}_{b-1}^T$,
	since otherwise it will be possible to add a scalar multiple of $H^{(0)}_l$ for some $l\in[1:b-1]$ to $H^{(0)}_0$ to get a vector $\un{v}=[v_0,v_1,\dots,v_{b}]^T$ such that $v_0=0,$ and such a vector lies in the span of $\{H^{(0)}_j\ |\ j\in[n-2b:n-b]\setminus\{n-2b\}\}$. Similarly, considering the $(2,b)$-burst whose support is given by $[0:b-1]\bigcup[n-2b:n-b]\setminus\{n-b\}$, we get $H^{(0)}(b,0)\ne0$ and $H^{(0)}(b,[1:b-1])=\un{0}_{b-1}^T$. 
	%We shall use this fact later in the proof.
	
	Now consider recoverability of the symbol $c_1$. Consider the $(2,b)$-burst whose support is given by $[1:b]\bigcup[n-2b+1:n-b+1]\setminus\{n-2b+1\}$. Following a similar argument as above, and noting that $H^{(1)}(0,[1:b-1])=\un{0}_{b-1}^T$, we get $H^{(1)}(1,1)\ne0$, and, if $b>2$, $H^{(1)}(1,[2:b-1])=\un{0}_{b-2}^T$. Noting that $H^{(0)}([0:b-1],[0:b-1])=H^{(1)}([0:b-1],[0:b-1]),$ we get $H^{(0)}([0:1],[0:1])$ is lower-triangular, with $H^{(0)}(i,i)\ne0,~i\in[0:1].$ Proceeding similarly, we consider recoverability of $c_2,c_3,\dots,c_{b-1}$ to get that $H^{(0)}([0:b-1],[0:b-1])$ is lower triangular, with $H^{(0)}(i,i)\ne0,~i\in[0:b-1]$. 
	%This in particular implies that $H^{(0)}([1:B-1],[1:B-1])$ is lower triangular, with $H^{(0)}(i,i)\ne0,~i\in[1:B-1]$.
	
	Now, we again focus on recovery of $c_0$.
	\barg
	Suppose there exists an index $s\in[b:n-3b+1]$ such that $H^{(0)}(\{0,b\},[s:s+L-1])$ has rank $2$, where $L\le b$. 
	Then, $H^{(0)}(\{0,b\},0)\in\text{span}\langle\{H^{(0)}(\{0,b\},j)\ |\ j\in[s:s+L-1]\}\rangle$.
	Since $H^{(0)}(0,[1:b-1])=\un{0}^T_{b-1},$ $H^{(0)}(b,[1:b-1])=\un{0}^T_{b-1}$ and $H^{(0)}([1:b-1],[1:b-1])$ is lower triangular with $H^{(0)}(i,i)\ne0,~i\in[1:b-1]$, we get that $H^{(0)}_0\in\text{span}\langle\{H^{(0)}_j\ |\ j\in[1:b-1]\bigcup[s:s+L-1]\}\rangle$. The erasure pattern whose support is given by $[0:b-1]\bigcup[s:s+L-1]$ is a $(2,b)$-burst, hence no such $s$ should exist. 
	\earg
	We will now show that such an $s$ exists by invoking Lemma \ref{lem:main}. 
	Since $H^{(0)}(0,[1:b-1])=\un{0}_{b-1}^T$, from Lemma \ref{lem:full_rank}, it follows that $H^{(0)}(0,b)\ne0$. Similarly, $H^{(0)}(b,b)\ne0$. If $u\ge2$, from Argument 1 as well as by noting that $\cc$ is an $[n=k+zb,k]$ systematic block code that is delay-$\taust$ decodable for any $(2,b)$-burst, it follows that $H^{(0)}(\{0,b\},[1:ub])$ satisfies all the properties stated in Lemma \ref{lem:main}, and thus $H^{(0)}(b,ub)\ne0$. Thus, $H^{(0)}(b,ub)\ne0$ for $u\ge 1.$
	Observing that $H^{(0)}(\{0,b\},n-2b)=[1,0]^T$, we get $H^{(0)}(\{0,b\},\{ub,n-2b\})$ is full-rank. 
	The burst erasure whose support is given by $[ub:n-2b]$ is of length $\le b$.
	%Since $H^{(0)}([1:B-1],[1:B-1])$ is lower-triangular with $H^{(0)}(i,i)\ne0,~i\in[1:B-1]$, we get $H^{(0)}_0\in\text{span}\{H^{(0)}_j:j\in[1:B-1]\bigcup[mB:k]\}$. 
	% yields an admissible erasure pattern. Thus, there exists an admissible erasure pattern from which code symbol $c_0$ is not recoverable within delay $T^\star$.
	
	\paragraph{Case $z>2$}
	Consider the code $\tilde{\cc}$ obtained by puncturing $\cc$ on the coordinates $\mathcal{P}=[n-zb:n-2b-1]$, i.e., the code $\tilde{\cc}$ is obtained by deleting all the coordinates in $\mathcal{P}$ from every codeword of $\cc$. Then, $\tilde{\cc}$ is a linear block code of dimension $\tilde{k}=k,$ and length $\tilde{n}=k+2b$. Let $\un{\tilde{c}}^T=[\tilde{c}_0,\tilde{c}_1,\dots,\tilde{c}_{\tilde{n}-1}]$ be any arbitrary codeword of $\tilde{\cc}$. Define a map $f$ from $[0:\tilde{n}-1]$ to $[0:n-1]$ as follows: 
	\beq
	f(i)=\begin{cases}
		i,~\text{if}~i\in[0:\tilde{k}-1],\\
		(z-2)B+i, ~\text{if}~i\in[\tilde{k}:\tilde{n}-1].
	\end{cases}
	\eeq
	For a set $I=\{i_1,i_2,\dots,i_l\}\subseteq[0:\tilde{n}-1]$, let $f(I)$ denote the set $\{f(i_1),f(i_2),\dots,f(i_l)\}$. 
	Suppose that for some $i\in[0:\tilde{k}-1]$ there exists a $(2,b)$-burst $\un{e}$ whose support is given by $J\in[0:\tilde{n}-1],$ such that $\tilde{c}_i$ is not recoverable within delay $(\tilde{n}-b)$ when $\un{e}$ occurs. Then, with regard to $\cc$ it follows that $c_{f(i)}$ is not recoverable within delay $(n-b)$ when the $(z,b)$-burst $\un{e}'$ whose support is given by $f(J)\bigcup\mathcal{P}$ occurs. Since $\cc$ is delay-$(n-b)$ decodable for any $(z,b)$-burst, such an $i$ does not exist. Thus, $\tilde{\cc}$ is a causal $[\tilde{n}=\tilde{k}+2b,\tilde{k}]$ block code that is delay-$(\tilde{n}-b)$ decodable for any $(2,b)$-burst, where $\tilde{k}\ge b,$ which as we have shown in the previous case can not exist.
	\eprf
	\begin{remark}
		The proof presented here shows a stronger result than the statement of the theorem, by showing that the $0$-th code symbol itself can not be decoded within delay $\taust$ if $b\nmid \taust$.
	\end{remark}

\end{document}